\documentclass[letterpaper, 10 pt, conference]{ieeeconf}  
\IEEEoverridecommandlockouts
\usepackage{cite}
\usepackage{amsmath,amssymb,amsfonts}
\usepackage{graphicx}
\usepackage{textcomp}
\usepackage{xcolor}
\usepackage{accents}
\usepackage[mathscr]{euscript}
\usepackage{mathrsfs}
\usepackage{mathtools}
\usepackage{textcomp, gensymb}
\usepackage{booktabs}
\usepackage{multirow}
\usepackage{makecell}
\usepackage{array}
\usepackage{algorithm, algpseudocode}
\usepackage{subcaption}
\usepackage[capitalize]{cleveref}
\usepackage{comment}

\begin{document}

\title{Koopman-Based Methods for EV Climate Dynamics: Comparing eDMD Approaches}

\author{Luca Meda~and~Stephanie~Stockar%
\thanks{Luca Meda (meda.10@osu.edu) and Stephanie Stockar (stockar.1@osu.edu) are with the  Department of Mechanical and Aerospace Engineering and the Center for Automotive Research, The Ohio State University, 930 Kinnear Road, Columbus, OH 43212 USA}}

\maketitle

\begin{abstract}
In this paper, data-driven algorithms based on Koopman Operator Theory are applied to identify and predict the nonlinear dynamics of a vapor compression system and cabin temperature in a light-duty electric vehicle. By leveraging a high-fidelity nonlinear HVAC model, the system’s behavior is captured in a lifted higher-dimensional state space, enabling a linear representation. A comparative analysis of three Koopman-based system identification approaches—polynomial libraries, radial basis functions (RBF), and neural network-based dictionary learning—is conducted. Accurate prediction of power consumption over entire driving cycles is demonstrated by incorporating power as a measurable output within the Koopman framework. The performance of each method is rigorously evaluated through simulations under various driving cycles and ambient conditions, highlighting their potential for real-time prediction and control in energy-efficient vehicle climate management. This study offers a scalable, data-driven methodology that can be extended to other complex nonlinear systems.
\end{abstract}

\section{Introduction}
In battery electric vehicles (BEV), effective thermal management of the heating, ventilation and air conditioning (HVAC) and cabin systems is crucial, as it directly influences the vehicle driving range \cite{haoSeasonalEffectsElectric2020}. However, the equations governing the temperature and pressure dynamics in HVAC systems are highly nonlinear, posing significant challenges for control design and real-time implementation  \cite{kibalama2021model}.

For this reason, developing fast-computing models capable of predicting the nonlinear dynamics of the vehicle HVAC system, ideally through linear representation, is of critical importance, as it would enable the design of optimal and real-time implementable controllers. In recent years, Koopman Operator Theory has gained increased attention because it offers a theoretically sound framework for capturing nonlinear system dynamics through the linear evolution of observable functions in a higher dimensional lifted space \cite{brunton2021modern}. \par The Koopman Operator is a linear but infinite-dimensional operator \cite{koopmanHamiltonianSystemsTransformation1931}. Therefore, numerical approximation methods and data-driven techniques have been studied to approximate it for practical modeling and control problems \cite{bakker2020koopman}. Extended Dynamic Mode Decomposition (eDMD) is one of the most widely used approaches for approximating the Koopman Operator on a finite-dimensional subspace spanned by a library of observable functions \cite{williams2015data}. The quality of the approximation and the convergence properties of eDMD algorithms significantly depend on the choice of the function dictionary. The dictionary can be predetermined based on some prior knowledge of the physics of the system, or trained and learned from collected data, using machine learning techniques and neural network architectures \cite{bakker2020koopman}. 
While these numerical approximation techniques have often been tested on simple, well-studied problems with available analytical solutions, their effectiveness in practical applications involving fully unknown dynamics, multiple states, inputs, and disturbances remains underexplored. Furthermore, while advanced data-driven approaches like Koopman Operator Theory have been used for HVAC modeling \cite{pan2023koopman}, there is a limited focus on the impact of modeling choices involved with eDMD to the quality of the system linearization. \par 
This paper addresses these limitations by investigating the application of Koopman Operator Theory  to a high-fidelity model of the vapor compression system integrated with a reduced-order cabin model. This model is employed to generate the data used for training the Koopman algorithms presented in this work. A comparative analysis is conducted among three classes of eDMD approaches, namely polynomial libraries, radial basis functions (RBF) and neural network dictionary learning, designed on the controlled nonlinear AC system, demonstrating the state prediction capabilities of each method on trajectories generated by the high-fidelity nonlinear physics-based model. Finally, the efficacy of the Koopman-based model is evaluated through a comparison against simulations of the high-fidelity nonlinear model under realistic driving cycles and external conditions. A key novelty of this work lies in the accurate prediction of power consumption over entire driving cycles, achieved by integrating power as a measurable output in the Koopman framework. The proposed methodology offers a scalable, data-driven solution that can be readily extended to other complex nonlinear systems, providing a pathway for the development of real-time implementable controllers for energy-efficient vehicle climate management.

\section{Methodology}
\subsection{Vapor Compression System and Cabin Modeling}\label{sec:theory}
\begin{figure}[b]
    \centering
        \centering
        \includegraphics[width=0.42\textwidth]{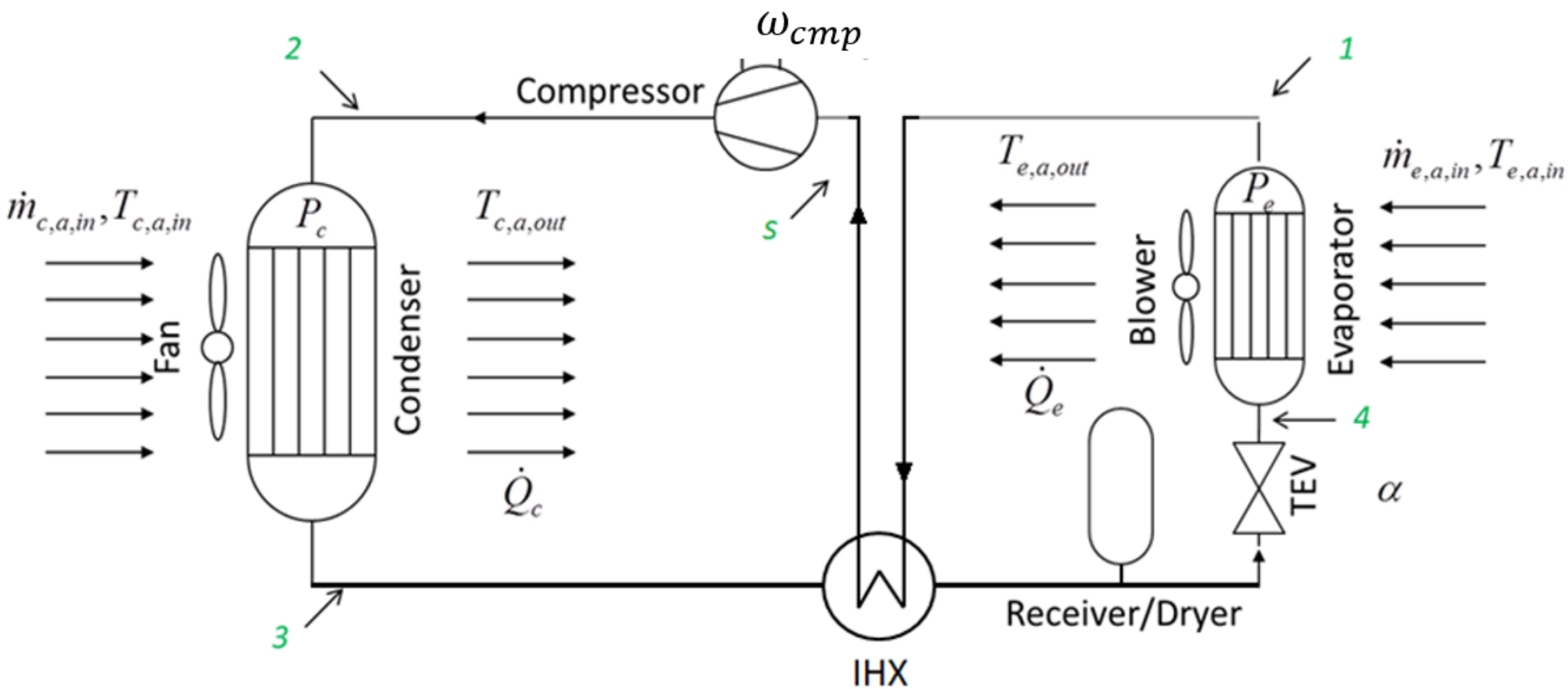}

        Vapor compression circuit diagram
        
        \label{fig:kp_AC_diagram}

        \centering
\includegraphics[width=0.42\textwidth]{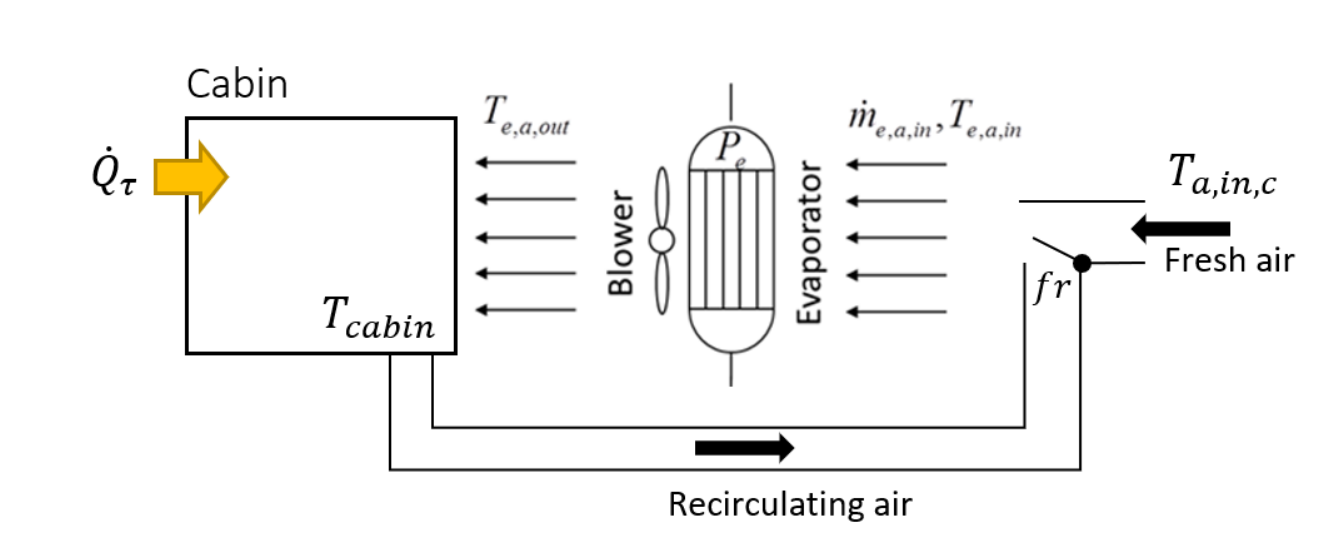}

        Cabin Model 
        \label{fig:kp_AC_cabin}

    \caption{Vehicle air conditioning and cabin model}
    \label{fig:cabin_model}
\end{figure}

The high-fidelity model considered in this paper consists of a lumped-parameter model of the vehicle vapor compression system integrated with a reduced-order cabin model as illustrated in Fig.~\ref{fig:cabin_model}, \cite{de2024optimization}.
The model captures the dynamics of the refrigeration loop, including the heat exchangers, compressor and expansion valve, and predicts the total energy consumption of the system during cabin heat removal. The refrigerant dynamics in the heat exchangers, as described in more details in prior work \cite{kibalama2021model}, are modeled using energy and mass conservation laws, along a set of assumptions that reduce the system representation to two states, the pressure in the evaporator $p_e$ and the pressure in the condenser $p_c$. A third state, the cabin temperature $T_{cabin}$, is then included to capture the interaction between the heat exchangers and the cabin system:

\begin{equation}
\label{eq:model}
\begin{aligned}
\left[\begin{array}{ccc}
d_{11}\left(p_e\right) & 0 & 0\\
0 & d_{22}\left(p_c\right) & 0 \\
0 & 0                      & d_{33}(T_{cabin})
\end{array}\right] \frac{d}{d t}\left[\begin{array}{l}
p_e \\
p_c \\
T_{cabin}
\end{array}\right]= \\
\left[\begin{array}{c}
\dot{Q}_e+\dot{m}_r\left(h_4-h_g\left(p_e\right)\right) \\
-\dot{Q}_c+\dot{m}_r\left(h_2-h_l\left(p_c\right)\right) \\
-\Dot{Q}_e+\Dot{Q}_{in}(T_{cabin})+\Dot{Q}_{gen}
\end{array}\right]
\end{aligned}
\end{equation}
where ${\dot{m}_r}$ is the refrigerant flow rate in the loop, $d_{11}(p_e)$ and $d_{22}(p_c)$ are function of the thermodynamic proprieties, the heat transfer rate in the heat exchargers is represented by $\Dot{Q}_e$ and $\dot{Q}_c$ for the evaporator and condenser respectively, the enthalpy change across the heat exchangers is given by $h_4-h_g$ and $h_2-h_l$, and \(\dot{Q}_{in}\), \(\dot{Q}_{gen}\) are the heat transfer rates respectively between the vehicle structure and the cabin, and between the vehicle structure and the ambient.\par 
The overall system energy utilization considers   power of the fan $P_{fan}$ and the power of the compressor $P_{cmp}$:
\begin{equation}
\label{eq:kp_Power_fan_cmp}
P_{cmp} = \frac{\dot{m}_r(h_2-h_l)}{\eta_{e,cmp}}
\,\,\,\,\
P_{fan} = \frac{1}{\eta_{e,fan}}\frac{P_{max}}{\omega^3_{max}}\omega^3_{fan}
\end{equation}
where $\eta_{e,cmp}$ and $\eta_{e,fan}$ are the compressor and fan electric motor efficiencies. The fan flow rate \(\dot{m}_{\text{fan}}\) and the speed of the compressor \(\omega_{\text{cmp}}\) are the inputs to the model while blower speed $\omega_{blw}$, the vehicle velocity $v_{veh}$ and the ambient temperature $T_{ac,in}$ are considered  disturbances. Finally, the cabin temperature is obtained by applying conservation of energy to the control volume, and assuming the cabin external mass exchanges heat with the outside environment through convection and solar absorption, and the cabin internal mass (representing components like carpets, seats) absorbs transmitted solar radiation and exchanges heat with the external mass through a surface with a defined thermal resistance \cite{de2024optimization,kwak2023thermal}.  

\subsection{Koopman Operator Theory}
Koopman Operator Theory, originally developed for autonomous systems, has been extended to incorporate control and exogenous inputs, allowing the operator to evolve over an augmented state space obtained as the product of the original state space and the space of input sequences \cite{kordaLinearPredictorsNonlinear2018}. Let us consider the generic discrete time dynamic system:
\begin{equation}
\label{eq:kp_dynamic_system}
\\\ x_{k+1}=f(x_{k},u_{k},w_k)
\end{equation}
where k $\in \mathbb{N}$ represents the time index, $f$ denotes the mapping forward in time, $x \in \mathbb{R}^n$ is the n-dimensional state vector, $u \in \mathbb{R}^{n_u}$ and $w \in \mathbb{R}^{n_w}$ are the input and disturbances vectors, respectively. The Koopman operator can represent the evolution of a dynamic system through its observables, defined as square-integrable functions ($\Psi: \mathbb{R}^n \rightarrow{} \mathbb{R}$):

\begin{equation}
    \label{eq:kp_operator}
\mathit{K}_{\Delta t} \Psi(x_k)  = \Psi\left(x_{k+1}\right)  =  \Psi(f(x_k,u_k,w_k))
\end{equation}

\subsection{Data-driven Koopman-based Modeling}
\label{model}
A computational representation of the infinite-dimensional operator is theoretically intractable. However, an approximation with a finite set of dictionary functions can yield good results for practical applications \cite{bakker2020koopman}. In this context, eDMD is one of the most popular approaches for obtaining a finite-dimensional approximation of the Koopman Operator based on measurements data of the system at hand. The idea behind the algorithm is to collect snapshot matrices $X$ and $X^+ \in \mathbb{R}^{n\times(T-1)}$, which contain states data collected for T-1 timestep and the corresponding states advanced by one timestep, respectively. These measurements can be obtained from different trajectories, as long as pairs of consecutive snapshots of the system are stacked in the matrices. The key step is to lift the dynamical system by applying a set of basis functions $\Psi_{i}$ to each $x_k \in X,X^+$, resulting in the lifted state vectors $Z,Z^+ \in $ $\mathbb{R}^{N\times(T-1)}$, where N is the dimension of the lifted state space (i.e. the number of basis functions chosen). The objective is then to fit a linear time invariant (LTI) matrix \(A\) and input and disturbances matrices \(B\) and \(D\) to the discrete-time system lifted in the higher-dimensional space defined by the basis function $\Psi_{i}$, resulting in the following LTI model:
\begin{equation}
    \begin{aligned}
       \label{eq:kp_lifted_dynamic_system}
    z_{k+1} & \approx A z_k+B u_k + D w_k\\
    \hat{x}_{k+1} & \approx C z_{k+1}
    \end{aligned}
\end{equation}
where $C$ is the matrix used to project back to the original state space, yielding the predicted state $\hat{x}$ at the subsequent timestep. The equality in the above relation holds only approximately if the choice of the basis of observables does not exactly span an invariant Koopman subspace \cite{haseli2022temporal}. A least-square problem is solved to find the best-fit linear operators given the data: 
\begin{equation*}
\underset{A,B,D}{argmin} \left\| Z^{+} - AZ - BU - DW \right\|_{F}
\end{equation*}

\begin{equation}
       \label{eq:kp_fit_matrices}
    [A,B,D]  = Z^{+} \begin{bmatrix} Z
 \\ U \\ W

\end{bmatrix}^{\dagger}
\end{equation}
where $\left\| . \right\|_{F}$ is the Frobenius norm, $\dagger$ indicates the Moore-Penrose pseudoinverse of the matrix. Since the goal is to predict the evolution of the system states, the states are included in the library of lifting functions, resulting in:
\begin{equation}
\label{states_lifted}
Z = \Psi(X) = [x, \psi_{1}(x), ......, \psi_{N-n}(x)]
\end{equation}
and consequently:
\begin{equation}
\label{projection}
x^{k+1} \approx Cz^{k+1} \; \text{with} \; C = [I_{n}, 0_{n \times (N-n)}]
\end{equation}
where $I_{n}$ is the identity matrix $(n\times n)$ and $0$ is the matrix with $n\times(N-n)$ null entries. 
To reduce the accumulation of errors arising from the approximation of the Koopman invariant subspace using  the library $\Psi(X)$, an additional correction step is incorporated in the process \cite{kordaLinearPredictorsNonlinear2018}, as described in~\cref{alg_2}. This modification of the standard approach results in an increased accuracy of the prediction, though it necessitates evaluating nonlinear functions on the states at each timestep. In practical applications where measurements from the system might be affected by sensor noise, a correction to eDMD can be introduced to reduce the induced bias by approximating Koopman matrices both forward and backward in time, as derived in \cite{dawson2016characterizing}.
\subsection{Selection of Dictionary of Basis Functions}\label{basis}
The choice of the library of basis functions is the most critical step for guaranteeing convergence of the proposed eDMD method and a good reconstruction accuracy \cite{brunton2021modern}. 
In this work, three different choices of dictionary of basis functions are implemented and compared,  with their interpretability and performance evaluated as the total number of functions varies. \par The first option consists of building a monomial-based library, containing all the polynomial combinations up to degree $m$ for each state, so that the dimension of the lifted state space (considering $N(0)=0$) is:
\begin{equation}
\label{linear_evolution}
N(m) = N(m-1) + \frac{(m+1)(m+2)}{2}
\end{equation}
This library is selected to reflect some of the known physics of the HVAC and cabin systems, and it is the simplest to compute and interpret. 
The computational time to obtain matrices A,B and D in~\eqref{eq:kp_fit_matrices} is $\mathbb{O}(N^2T)$ and $\mathbb{O}(N^3)$, considering that in our setting the two dimensions satisfy $T>>N$. 
\par 

The second library option explored in this paper extends the library of analytical functions to include RBF, which have shown good performance for eDMD  in various applications \cite{williams2015data}. The classes of RBF chosen are thin plate spline \eqref{thin}, gaussian \eqref{gaussian}, and inverse quadratic \eqref{inverse}: 
\begin{subequations}
\label{rbf}
\begin{align}
\label{thin}
  \Psi(x) &= \left\|x-x_c \right\|^2 log \left\|x-x_c \right\|  \\
  \label{gaussian}
  \Psi(x) &= exp(-\epsilon^2\left\|x-x_c \right\|^2)
\\
\label{inverse}
  \Psi(x) &= \frac{1}{\sqrt[]{1+\epsilon^2}\left\|x-x_c \right\|^2}
\end{align}
\end{subequations}

When applying RBFs to each snapshot $x_k$, the resulting dimension of the lifted space $N$ depends on the dimension of the vector of centers $x_c$. The K-means algorithm is used  to identify $N-3$, where 3 is the number of states,  different clusters in the data, with each cluster assigned a  center $x_{c,i}$. 
\par

The final method introduced to derive a candidate library of lifting functions aims to eliminate the need to manually select a set of functions by introducing a parametric dictionary $\Psi(X) = \Psi(X,\theta)$ via a neural network trained to learn a dictionary given the available data and a loss function to minimize. The method, first introduced as eDMD with Dictionary Learning for autonomous systems (eDMD-DL)\cite{li2017extended} and extended here to a controlled system (\cref{alg_1}), offers the advantage of not requiring prior knowledge of the system. However, this comes at the expense of increased training resources and a lack of  control over the functions used to span the approximate Koopman subspace. The algorithm requires the dimension of the trainable output of the network \((N-3)\) to be fixed, with the state matrix $X$ serving as the input and $X^+$ set as the non-trainable output. This ensures that the states are included in the resulting dictionary, as in~\eqref{states_lifted}. It is noted that the vectors $u$ and $w$ are left unlifted to facilitate the design of linear control strategies based on the Koopman model. A one-step prediction error loss function is used to optimize the vector parameter $\theta$, which contains the weights and biases of the network. This approach is particularly effective for this application, where the objective is short-term prediction of a limited number of observables.

\begin{algorithm}[b]
    \caption{eDMD-DL with inputs}
    \label{alg_1}
    \begin{algorithmic}[1]
        \State{\text{Collect Data}: $\left\{x_{k+1},x_{k},u_k,w_k\right\}_{k=0}^{T-1}$}
        \State{\text{Initialize random} $\theta,K$}
        \State{\text{Set network hyperparameters}: $\delta > 0$ \text{learning rate}, $\epsilon > 0$ \text{tolerance}, $Opt$ \text{Optimizer}, $L=L(K,\theta)$ \text{loss}}
        \State{$L = \sum_{k=0}^{T-1}\left\| \Psi(x_{k+1},\theta) -K[\Psi(x_k,\theta);u_k;w_k])\right\|^2$}
        \While{$L(K,\theta)>\epsilon$}
        \State{\text{Evaluate} $L(K,\theta)$}
        \State{$\theta \gets \theta - \delta\bigtriangledown_{\delta}L(K,\theta)$ \text{according to $Opt$}}
        \EndWhile
        \State{\text{Compute} $Z = \Psi(X,\theta)$, $Z^+ = \Psi(X^+,\theta)$}
    \end{algorithmic}
\end{algorithm}
\section{Case Study}\label{sec:results}
\subsection{Problem Description}\label{sec:problems}
The nonlinear dynamics of the AC has three states $x_{i}$ ($p_e,p_c,T_{cabin}$), two control inputs $u_{i}$ (\(\dot{m}_{\text{fan}}\), \(\omega_{\text{cmp}}\)), and three exogenous inputs $w_{i}$ (\(T_{a,\text{in},c}\), \(v_{\text{veh}}\), \(\omega_{\text{blw}}\)). Furthermore, data for the two system outputs, the compressor power $P_{cmp}$ and the fan power $P_{fan}$ are generated using the high-fidelity model for each timestep, and they are collected and stored in the matrix $Y \in \mathbb{R}^{2\times(T-1)}$. The outputs are integrated into the Koopman modeling framework, resulting in a linear representation for $Y$ through the nonlinear lifting of the states represented by $Z$:
\begin{equation}
\underset{E}{argmin} \left\| Y - EZ \right\|_{F}
\end{equation}
The linear representation through the operator E is advantageous for managing 
the nonlinearities in the power consumption relation and predicting the energy consumption of the system.

\subsection{Data Generation}
Simulations of the nonlinear AC system and cabin model are performed to generate the dataset used in this paper. The high-fidelity model is implemented in MATLAB/Simulink with a time resolution of 1s. A total of 200 different trajectories are generated using input sequences generated from pseudo-random truncated normal distributions, with sampled values constrained within the physical boundaries reported in~\cref{table_1}. 
The resulting data is then divided into training and test set depending on the length of the simulation. Trajectories lasting 8500s are collected in the training set, while trajectories of 1500s duration are used to validate the accuracy of the models.
\begin{table}[b]
\caption{Validity ranges for input and disturbances data generation}
\label{table_1}
\centering
\begin{tabular}{cccc}
\toprule
Input parameter & Unit of measure & Min & Max \\
\midrule
$\dot{m}_{fan}$ & kg/s & 0.01 & 0.48 \\
$\omega_{cmp}$ & $Hz$ & 13 & 83\\
$T_{a,in,c}$ & \textdegree C & 26 & 34 \\
$v_{veh}$ & km/h & 0 & 80 \\
$\omega_{blw}$ & m/s & 0.8 & 1.6 \\
\bottomrule
\end{tabular}
\end{table}
Each simulation is initialized with different initial conditions, and the inputs are varied randomly at 60-second intervals to excite the system's response. The data generation process is designed to achieve broad coverage of the state space while capturing the system response to different initial conditions, control inputs and external disturbances. The results of the training simulation are then stored in matrices $X,X^+,U,W,Y$ to train the Koopman models.

\subsection{Comparison of Libraries on Simulation Data}
\begin{figure}[b]
    \centering
    \includegraphics[width=1\linewidth]{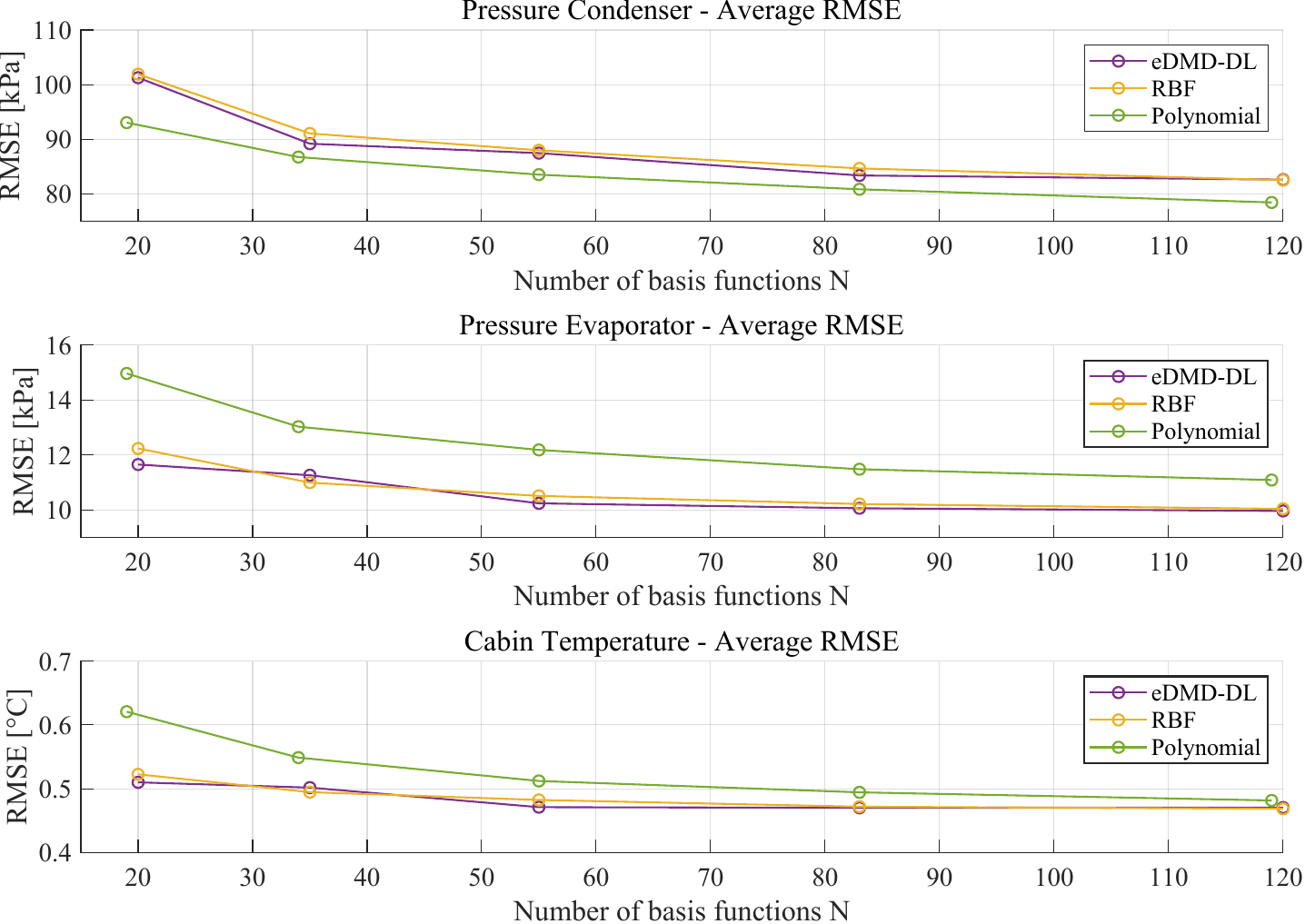}
    \caption{$RMSE_{avg}$ for the three system states compared between different dictionary choices with varying N}
    \label{fig_rmse}
\end{figure}

The three methods to derive a set of lifting functions are trained and tested using the data generated from simulations. The polynomial library is tested for degrees up to $m=7$, resulting in lifted space dimensions ranging from $N=19$ to $N=119$. The RBF dictionary class is applied to the problem with dimension N ranging from 20 to 120 functions and the associated hyperparameters are optimized using the GridSearchCV algorithm on the test set, yielding optimal values of   $\epsilon=1$, see ~\cref{gaussian}.
The hyperparameters used in the implementation of the eDMD-DL procedure are provided in~\cref{table_2}. The output dimension N for this algorithm is varied between N=20 to N=120.
\begin{table}[b]
\caption{Hyperparameters of the eDMD-DL with inputs algorithm}
\label{table_2}
\centering
\begin{tabular}{cccc}
\toprule
Input parameter & Value \\
\midrule
$\delta$ & $10^{-4}$ \\
$\epsilon$ & $10^{-3}$ \\
$Opt$ & Adam \\
$epochs$ & 50 \\
$\text{Activation function}$ & $tanh$ \\
$\text{Hidden layers}$ & 3 \\
$\text{Neurons per hidden layer}$ & 2*N \\
\bottomrule
\end{tabular}
\end{table}
The representation of the Koopman model of \eqref{eq:kp_fit_matrices} is obtained by training the model with different choices of $\Psi(X)$ from the function classes and methods illustrated in~\cref{basis}, while varying the dimension N of the lifted state space for each method. The trained linear systems are then validated on the 200 test trajectories simulated from the nonlinear high-fidelity model, following the procedure outlined in~\cref{alg_2}. 
\begin{algorithm}[t]
    \caption{eDMD prediction with one-step correction}
    \label{alg_2}
    \begin{algorithmic}[1]
        \State{\text{Initialize states}: $x_0$. 
        \text{Collect input sequences:} $u_k,w_k$}
        \State{\text{Choose library $\Psi$ and dimension N}}
        \State{\text{Lift initial condition}: $z_0 = \Psi(x_0)$}
        \State{\text{Compute initial output}: $y_0 = Ez_0$}
        \For{$0 \le k <T$}
        \State{$z_{k+1} = Az_k + Bu_k + Dw_k$ }
        \State{$\hat{x}_{k+1} = Cz_{k+1}$}
        \State{$\hat{y}_{k+1} = Ez_{k+1}$}
        \State{$z_{k+1} \gets \Psi(\hat{x}_{k+1})$}
        \EndFor
        \State{\text{Store predicted states} $\left\{\hat{x}_k  \right\}_{k=0}^T $\text{ and outputs} $\left\{\hat{y}_k  \right\}_{k=0}^T$}
    \end{algorithmic}
\end{algorithm}
For each method and each dimension N,  an average RMSE metric is computed to evaluate and compare the prediction accuracies of the models on the test sequences of inputs and states:
\begin{equation}
\label{rmse}
RMSE_{avg} = \frac{1}{N_{sim}}\sum_{j=1}^{N_{sim}}\sqrt{\frac{{\sum_{_{K}}^{}\left\| \hat{x}^j(k)-x_{true}^j(k)) \right\|_{2}^{2}}}{k}}
\end{equation}
where $N_{sim}=200$ is the total number of test simulations, and $k$ is the time index for each simulation. The same metric is computed for the power consumption predicted values $\hat{y}_k$.
In Fig.~\ref{fig_rmse}, the $RMSE_{avg}$ is shown for the three classes of dictionary as the dimension N varies. It was found that the thin plate spline RBF achieves that best accuracy among the four functions defined in~\eqref{rbf}, so this RBF library is compared against the polynomial one and the eDMD-DL results. 
It was observed that, for the condenser pressure, the polynomial library provides the best accuracy for any value of N. However, for the evaporator pressure and cabin temperature, the polynomial library exhibits low accuracy at small \(N\). As N increases, the accuracy gap between the polynomial library, the RBF, and eDMD-DL stabilizes at $2.5\%$ for the cabin temperature and $9.4\%$ for the evaporator pressure. The RBF library shows an accuracy comparable to the eDMD-DL dictionary, indicating that the use of a neural network-based dictionary learning process does not offer significant improvements in identifying the states of the AC system, but additional training effort is required. \par
The Consistency Index (CI) was calculated to assess the accuracy of eDMD in representing the dynamics of a system within a finite subspace spanned by a dictionary of functions, as introduced in \cite{haseli2022temporal}. This metric quantifies how closely the chosen set of observables approximates an invariant Koopman subspace. A lower CI value indicates that the subspace is more consistent with the true dynamics of the system, meaning it better captures the system's evolution over time. The results show that the polynomial and RBF subspaces achieve lower CI values compared to the eDMD-DL libraries, suggesting that they are closer to being Koopman invariant. From the analysis of the CI index and the relationship of $RMSE_{avg}$ versus the dimension N for polynomial and RBF libraries (Figure \ref{fig_rmse}), it is observed that increasing the dimension of the lifted space beyond $N>=35$ does not significantly improve model accuracy. This suggests that a more parsimonious library may be sufficient to represent the original nonlinear system and avoid the risk of overfitting. Furthermore, both the polynomial and RBF libraries are based on physics-derived analytical functions. 

\begin{figure}[b!]
    \centering
        \centering
        \includegraphics[width = 0.48\textwidth]{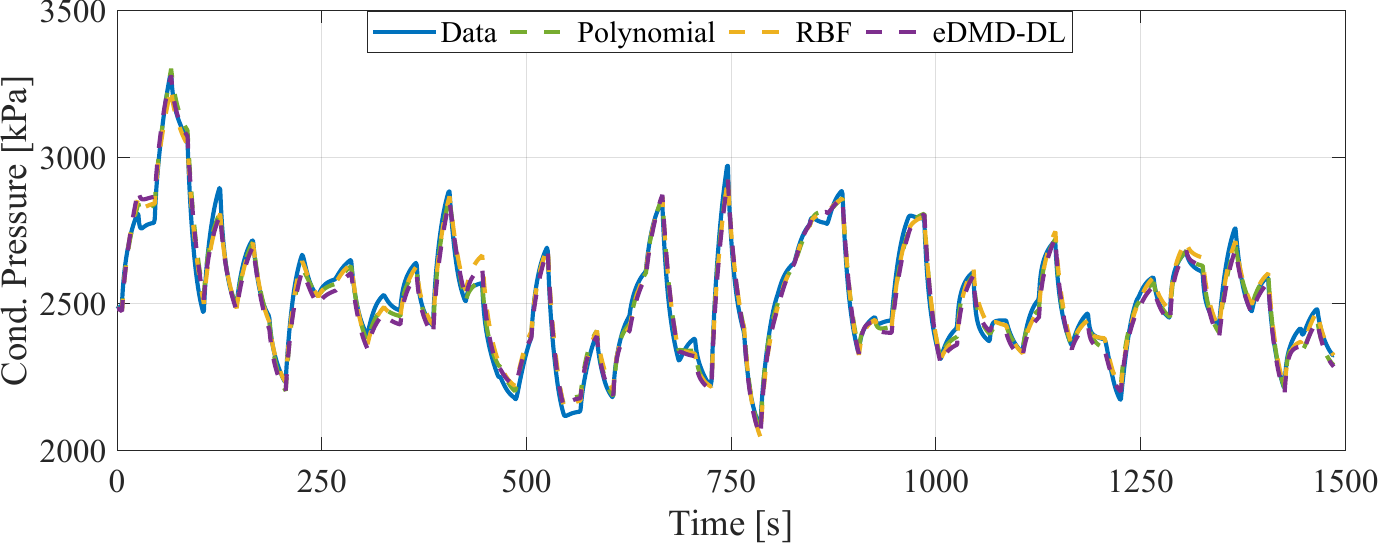}
        
        \centering
        \includegraphics[width = 0.48\textwidth]{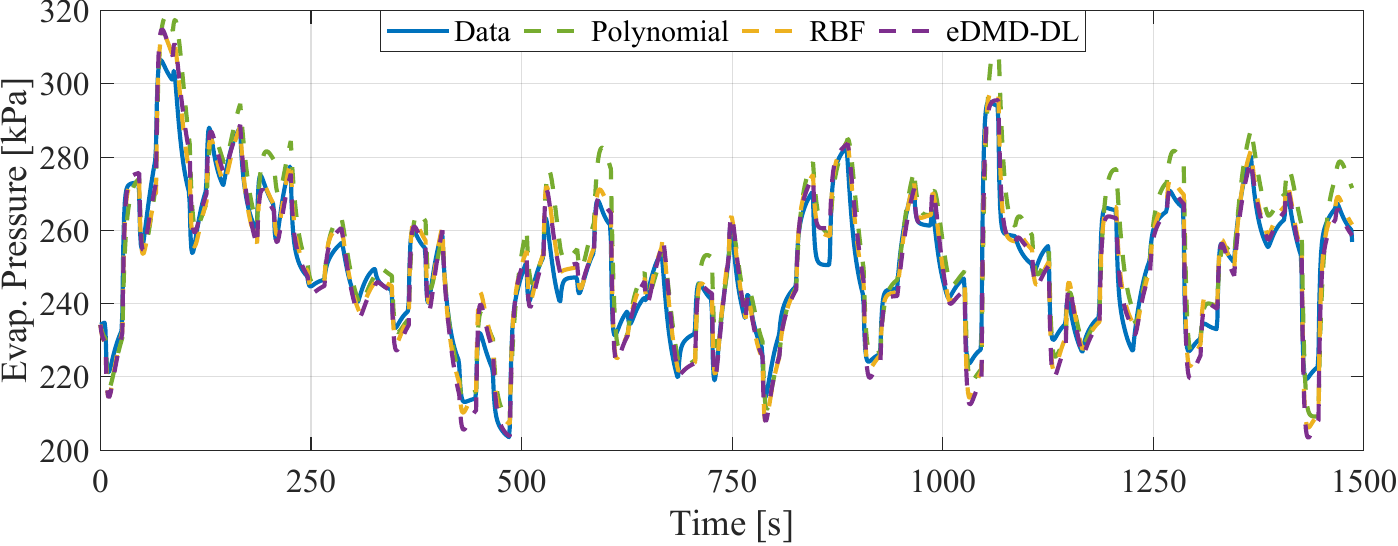}
        
        \centering
        \includegraphics[width = 0.48\textwidth]{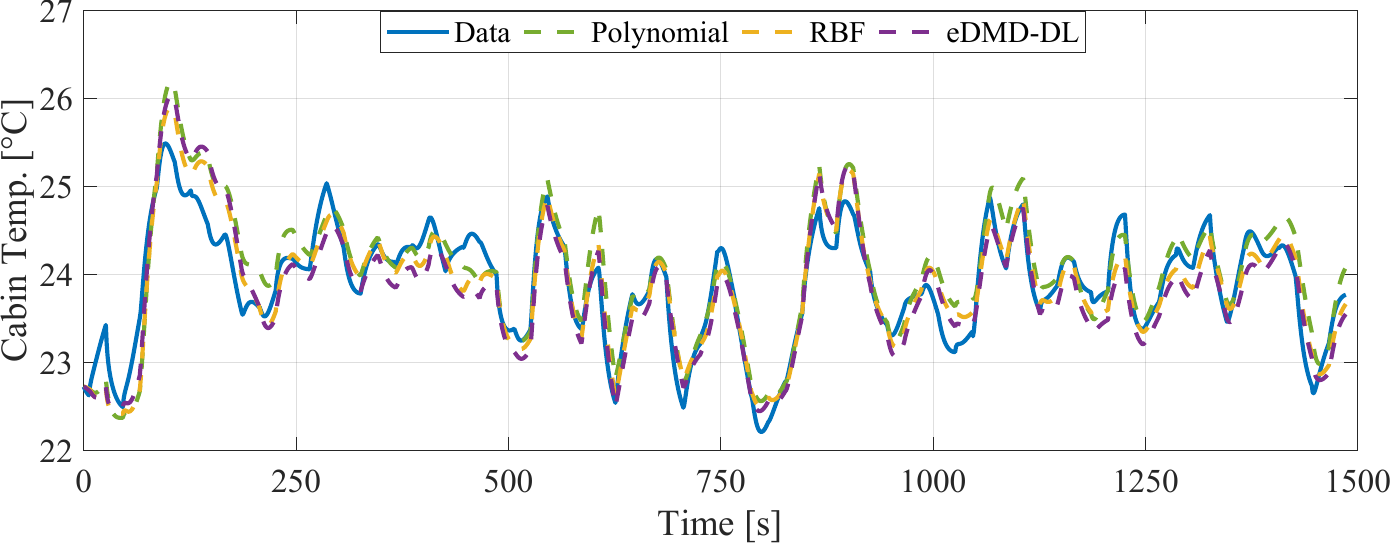}

        \centering
        \includegraphics[width = 0.48\textwidth]{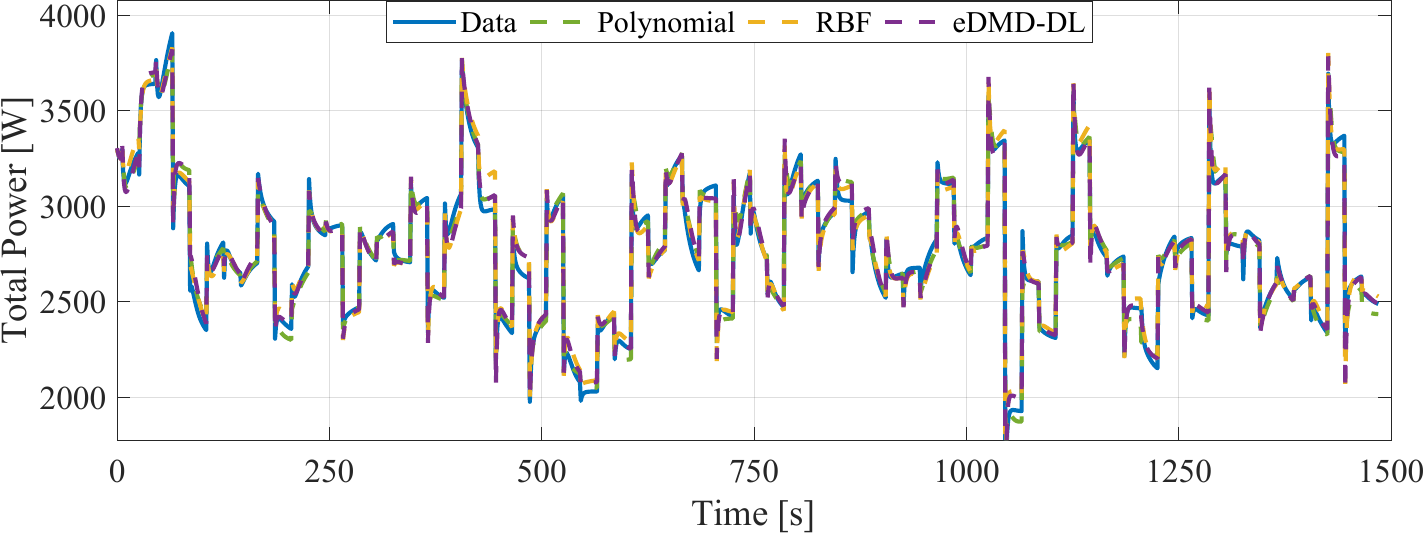}
        
\caption{States and power prediction for the Koopman methods tested (N=35) compared to a 1500s validation sequence}
\label{fig_test}
\end{figure}

Fig.~\ref{fig_test} shows a direct comparison between the three original states $\hat{x_{i}}$, and the total power $\hat{y_1} + \hat{y_2}$, with the states and outputs predicted using the three Koopman identification methods for \(N=35\) over one test. Predictions are made 1500s into the future starting from a known initial condition, representing a complete open-loop prediction. For practical  applications, the prediction horizon is typically  shorter, leading to lower $RMSE_{avg}$ values than those reported in this paper \cite{kibalama2021model}. Consequently, this simulation setup represents a worst-case scenario for prediction accuracy. 

\subsection{Driving scenarios results}\label{sec:driving_scenario}
In this section, the validation is extended beyond the 200 test sequences with randomly generated inputs by  defining a set of new inputs $u_{i}$ and external disturbances $w_{i}$ representative of the AC and cabin behaviour for a realistic driving cycle. Based on the results obtained in the previous section, this comparison focuses on the analytical dictionary. Specifically, the results are obtained using the thin plate spline RBF dictionary with $N=35$ as lifting library $\Psi$. \par
\begin{figure}[t]
      \centering
      \includegraphics[width=0.8\linewidth]{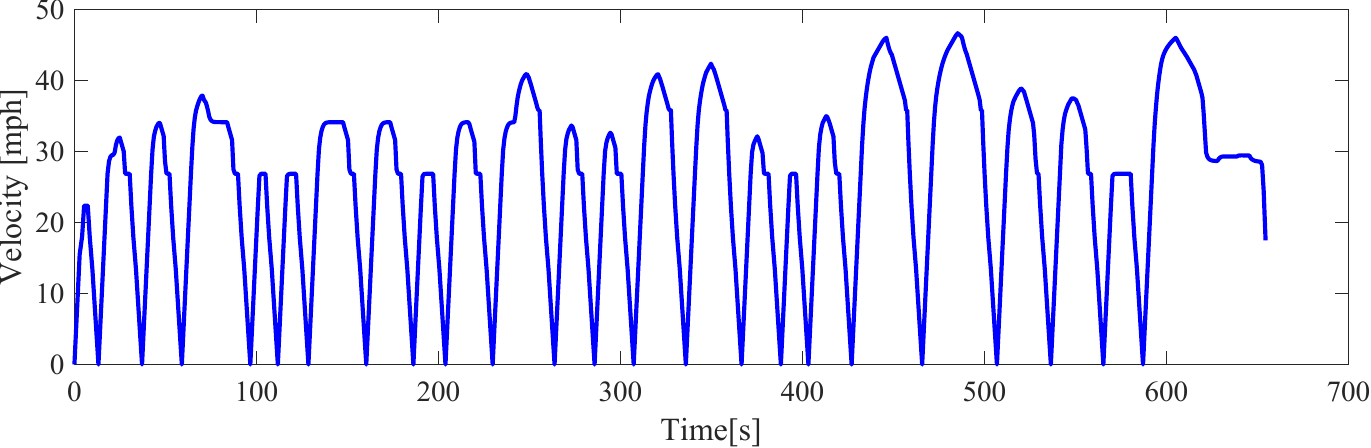}  
\caption{Route 15 velocity profile}
\label{fig_route_15}
\end{figure}
\begin{figure}[t!]
    \centering
        \centering
        \includegraphics[width = 0.43\textwidth]{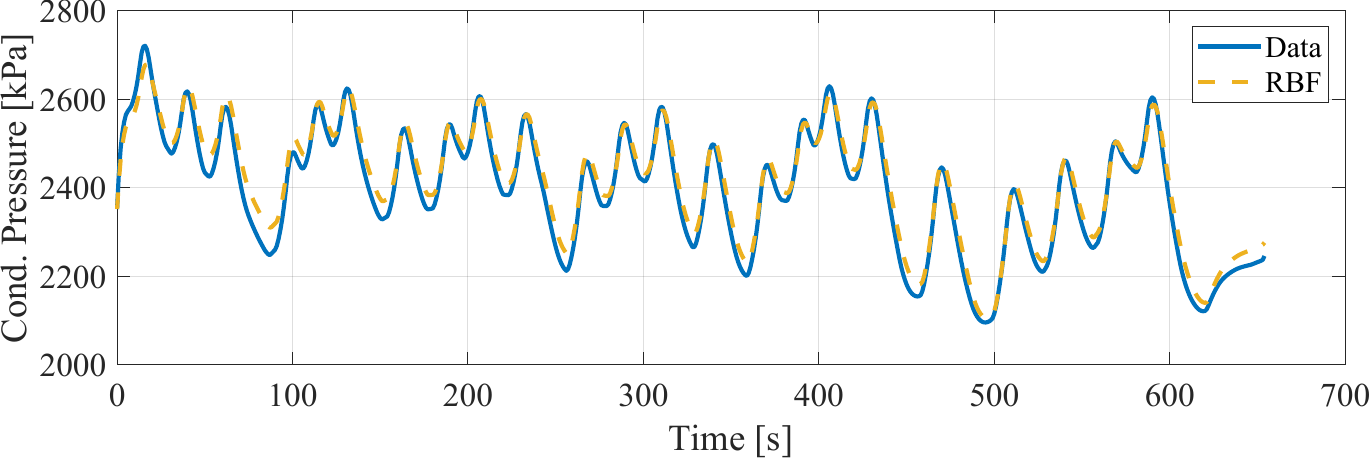}
        
        \centering
        \includegraphics[width = 0.43\textwidth]{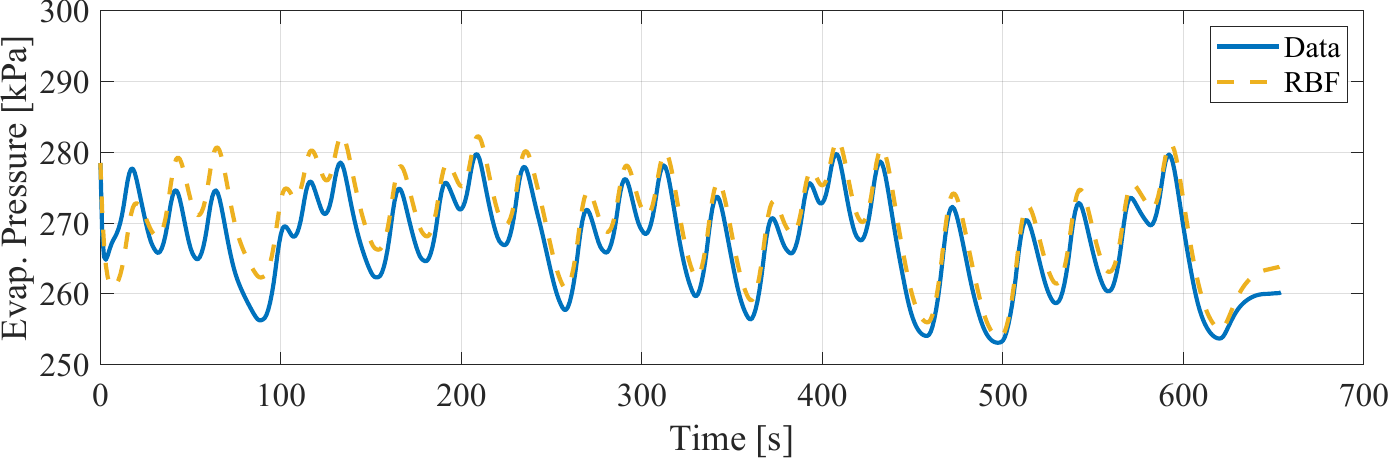}
        
        \centering
        \includegraphics[width = 0.43\textwidth]{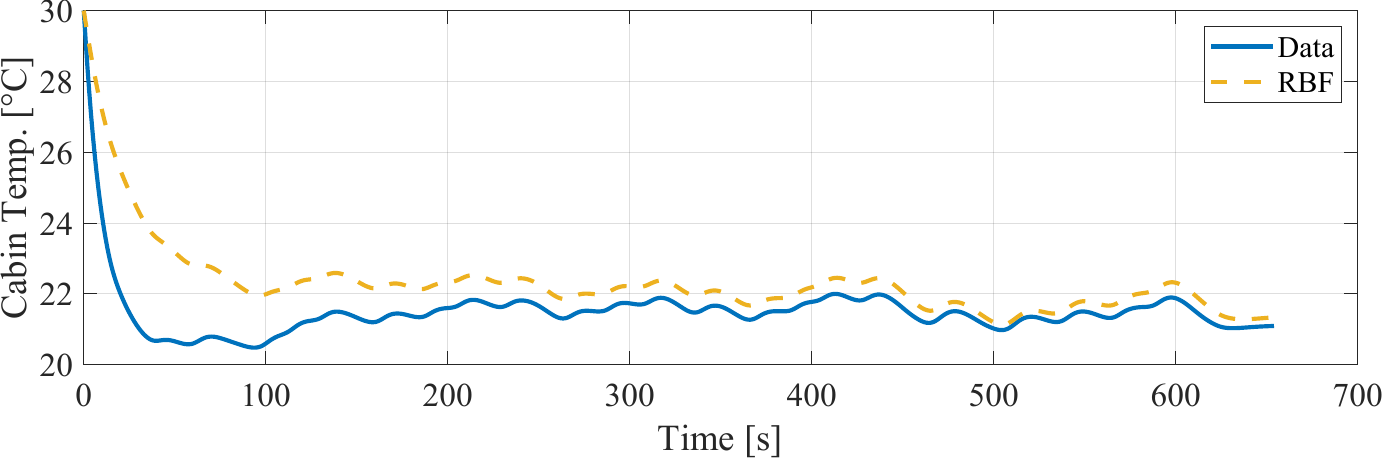}

        \centering
        \includegraphics[width = 0.43\textwidth]{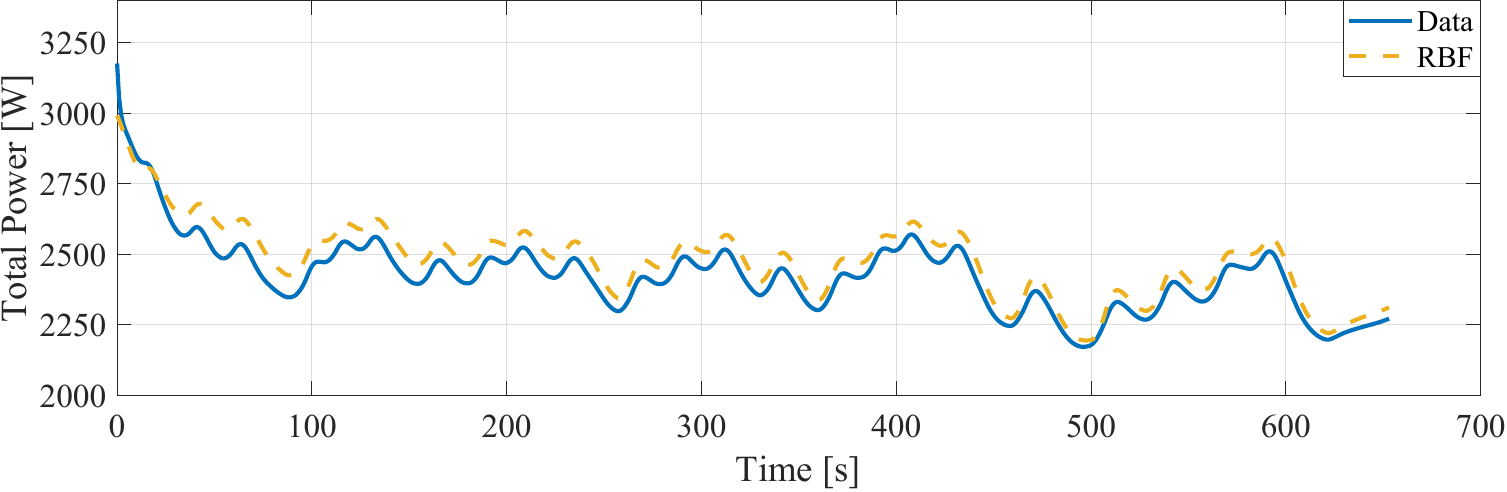}

\caption{States and power prediction for RBF Koopman method on the Route 15 driving cycle}
\label{fig_route_15_prediction}
\end{figure}

The velocity profile of Fig. \ref{fig_route_15} is used as external input $v_{veh}$. This cycle, Route 15, includes vehicle velocity data collected over a 7.5km route in Columbus, Ohio and it is characterized by frequent traffic lights and stop signs \cite{ozkan2024optimizing}. Additional disturbances, $\omega_{blw}$ and $T_{ac,in}$, are kept constant throughout the 650s simulation. Control inputs \(\dot{m}_{\text{fan}}\) and \(\omega_{\text{cmp}}\) are generated implementing a simple baseline controller on the nonlinear system designed to track a reference temperature $T_{cabin}$ \cite{de2024optimization}. 
Fig.~\ref{fig_route_15_prediction}  shows the results of a simulation performed on Route 15, with $T_{ac,in}=30^\circ C$, $\omega_{blw} = 1.2$ m/s. The other control inputs are obtained by implementing a cooling strategy on the high-fidelity nonlinear system to track a reference temperature of $23^\circ C$. The corresponding input sequences are then implemented in open-loop. \par
The RBF lifting library shows strong performance in predicting the three states and power consumption, even under realistic driving conditions. The largest prediction error occurs during the initial transient phase of $T_{cabin}$. Accurately predicting energy consumption over the entire driving cycle is crucial for AC systems.  For the RBF dictionary, the cumulative energy error is $2\%$. Table~\ref{tab_3} summarizes the prediction results of the RBF library on additional driving cycles, namely the SC03 and WLTP cycles, considering  variations in external inputs and external temperature.

\vspace{-5pt}

\begin{table}[t]
 \setlength{\tabcolsep}{5.5 pt}
    \centering
\begin{tabular}{ccc|cccc}
 \multicolumn{3}{c|}{\textbf{Parameters}}           & \multicolumn{4}{c}{\textbf{Percentage RMSE [\%]}}                                                                   \\ \hline
\multicolumn{1}{c|} {Cycle} & \multicolumn{1}{c|}{$T_{ac,in}[^\circ C]$} & $\omega_{blw}[m/s]$ & \multicolumn{1}{c|}{$p_c$} & \multicolumn{1}{c|}{$p_e$} & \multicolumn{1}{c|}{$T_{cabin}$} & Power \\ \hline
\multicolumn{1}{c|}{R15} & \multicolumn{1}{c|}{$30$} & 1.2 &
\multicolumn{1}{c|}{0.80} & \multicolumn{1}{c|}{1.04} & \multicolumn{1}{c|}{3.76} & 2.31 
\\
\multicolumn{1}{c|}{SC03} & \multicolumn{1}{c|}{$30$} & 1.2 &
\multicolumn{1}{c|}{0.55} & \multicolumn{1}{c|}{1.07} & \multicolumn{1}{c|}{3.80} & 2.14 \\
\multicolumn{1}{c|}{WLTP} & \multicolumn{1}{c|}{$30$} & 1.2 &
\multicolumn{1}{c|}{0.34} & \multicolumn{1}{c|}{0.15} & \multicolumn{1}{c|}{0.88} & 2.11
\\
\multicolumn{1}{c|}{R15} & \multicolumn{1}{c|}{$27$} & 1.0 &
\multicolumn{1}{c|}{1.40} & \multicolumn{1}{c|}{2.77} & \multicolumn{1}{c|}{5.40} & 0.97
\\
\multicolumn{1}{c|}{SC03} & \multicolumn{1}{c|}{$27$} & 1.0 &
\multicolumn{1}{c|}{0.69} & \multicolumn{1}{c|}{2.42} & \multicolumn{1}{c|}{5.45} & 1.03 \\
\multicolumn{1}{c|}{WLTP} & \multicolumn{1}{c|}{$27$} & 1.0 &
\multicolumn{1}{c|}{0.22} & \multicolumn{1}{c|}{2.33} & \multicolumn{1}{c|}{2.94} & 2.13 

\end{tabular}
    \caption{Parameters and prediction errors for driving cycle tests with the RBF lifting library (N=35)}
    \label{tab_3}
\end{table}

\section{Conclusions}\label{sec:conclusions}
      This paper presents a novel approach to model the nonlinear dynamics of  HVAC and cabin systems in a light duty vehicle 
      based on eDMD. The method constructs an approximation of Koopman Operator to provide a linear representation of the original system in a higher dimensional state space. Three different methods for lifting the system through libraries of nonlinear functions are compared. Results demonstrate that physics-based strategies, such as  those using polynomial functions and RBFs, outperform the black-box eDMD-DL algorithm for this specific application. 
      The most effective method, using RBF, was further validated through simulations representative of realistic driving conditions. The thin plate spline RBF-based library achieved the highest accuracy, effectively predicting state dynamics and  power consumption with only 35 lifting functions. This demonstrates the potential of the proposed method to provide a compact and efficient model representation. Future research will focus on developing linear controllers based on the identified Koopman approximation of the system, highlighting computational advantages and energy savings. 

\bibliographystyle{IEEEtran}
\bibliography{sources}

\end{document}